\documentclass[aps,showpacs,eqsecnum,10pt]{revtex4}
\usepackage{graphicx}
\usepackage{amsmath}
\usepackage{amssymb}

\newcommand{\be}{\begin{equation}}
\newcommand{\ee}{\end{equation}}
\newcommand{\ba}{\begin{eqnarray}}
\newcommand{\ea}{\end{eqnarray}}

\begin{document}


\title{Sum Rules for the Optical and Hall Conductivity in Graphene}

\author{V.P.~Gusynin$^{1}$}
\email{vgusynin@bitp.kiev.ua}
\author{S.G.~Sharapov$^{2}$}
\email{sharapov@bitp.kiev.ua}
\author{J.P.~Carbotte$^{2}$}
\email{carbotte@mcmaster.ca}

\affiliation{$^1$ Bogolyubov Institute for Theoretical Physics, 14-b
        Metrologicheskaya Str., Kiev, 03143, Ukraine\\
        $^2$ Department of Physics and Astronomy, McMaster University,
        Hamilton, Ontario, Canada, L8S 4M1}

\date{\today }

\begin{abstract}
Graphene has two atoms per unit cell with quasiparticles exhibiting
the Dirac-like behavior. These properties lead to  interband in
addition to intraband optical transitions and  modify the $f$-sum
rule on the longitudinal conductivity. The expected dependence of
the corresponding spectral weight on the applied gate voltage $V_g$
in a field effect graphene transistor is $\sim \mbox{const}-
|V_g|^{3/2}$. For $V_g =0$, its temperature dependence is $T^3$
rather than the usual $T^2$. For the Hall conductivity, the
corresponding spectral weight is determined by the Hall frequency
$\omega_H$ which is linear in the carrier imbalance density $\rho$,
and hence proportional to $V_g$, and is different from the cyclotron
frequency for Dirac quasiparticles.
\end{abstract}





\maketitle

\section{Introduction}
\label{sec:intro}

The real part of frequency dependent optical conductivity
$\sigma_{xx}(\Omega)$ is its absorptive part and its spectral weight
distribution as a function of energy ($\hbar \Omega$) is encoded
with information on the nature of the possible electronic
transitions resulting from the absorption of a photon. Even though
the relationship of the conductivity to the electronic structure and
transport lifetimes is not straightforward, much valuable
information can be obtained from such data. In particular, the
$f$-sum rule on the real part of $\sigma_{xx}(\Omega)$ stated in its
simplest form for an infinite free electron band,
\begin{equation}
\label{sum-rule-free} \int_{0}^{\infty}d \Omega \mbox{Re}
\sigma_{xx}(\Omega) = \frac{\omega_P^2}{8}
\end{equation}
has proved particularly useful. Here $\omega_P$ is the plasma
frequency, $\omega_P^2 =  4 \pi n e^2/m$, with $n$  the free carrier
density per unit volume, $-e<0$ the charge of electron, and $m$ its
effective mass. In this case the right hand side (RHS)  of
Eq.~(\ref{sum-rule-free}) is independent of temperature and of the
interactions. More generally for a finite tight binding band, the
optical sum rule takes the form
\be \label{sum-rule-1band} \frac{2}{\pi}\int_{0}^{\Omega_M} d\Omega
\mbox{Re} \, \sigma_{xx}(\Omega) =  \frac{e^2}{\hbar^2V}
\sum_{\mathbf{k},\sigma} n_{\mathbf{k},\sigma} \frac{\partial^2
\epsilon_\mathbf{k}}{\partial k_x^2}, \ee
with $\Omega_M$ a cutoff energy on the band of interest and only the
contribution to $\mbox{Re} \, \sigma_{xx}(\Omega)$ of this
particular band is to be included in the integral. Here $\hbar$ is
Planck's constant, $V$ is the crystal volume, $\sigma$ is the spin,
$\epsilon_\mathbf{k}$ is the electronic dispersion, $\mathbf{k}$ is
the wave vector in the Brillouin zone, and $n_{\mathbf{k},\sigma}$
is the probability of occupation of the state $|\mathbf{k}, \sigma
\rangle$. If we assume a free electron dispersion,
$\epsilon_\mathbf{k} = \hbar^2 \mathbf{k}^2/2m$,
Eq.~(\ref{sum-rule-1band}) immediately reduces to
Eq.~(\ref{sum-rule-free}) with $n=N/V$, where $N = \sum_{\mathbf{k},
\sigma} n_{\mathbf{k},\sigma}$ is the total number of electrons in
the band. For tight-binding dispersion with nearest neighbor hopping
on a square lattice it is easy to show that the RHS  of
Eq.~(\ref{sum-rule-1band}) reduces to $e^2/\hbar^2$ multiplied by
minus one-half of the kinetic energy, $W_{\mathrm{K.E.}}$ per atom.
In this particular case the optical sum rule can be used to probe
the change in kinetic energy of the electrons as a function of
temperature or with opening of a gap in the spectrum of the
quasiparticle excitations which has been a topic of much recent
research
\cite{Hirsch1992PC,Millis:book,Marel:book,Benfatto:review,Carbotte:review}.

When a constant external magnetic field $\mathbf{B}$ is applied to a
metallic system in the $z$-direction, the optical conductivity
acquires a transverse component $\sigma_{xy}(\Omega)$ in addition to
the longitudinal component $\sigma_{xx}(\Omega)$. This quantity
gives additional information on the electronic properties modified
by the magnetic field. In this case Drew and Coleman
\cite{Drew1997PRL} have derived a new sum rule on the optical Hall
angle $\theta_H(\Omega)$. If we define
\begin{equation}\label{t_H}
t_H(\Omega) \equiv \tan \theta_H(\Omega) =
\frac{\sigma_{xy}(\Omega)}{\sigma_{xx}(\Omega)},
\end{equation}
then
\begin{equation}
\label{sum-rule-Hall} \frac{2}{\pi} \int_0^\infty d \Omega
\mathrm{Re} t_H(\Omega) = \omega_H,
\end{equation}
where $\omega_H$ is the Hall frequency which corresponds to the
cyclotron frequency $\omega_c = eB/m c$ for free electrons. Here $c$
is the velocity of light.

Recently, graphene which is a single atomic layer of graphite, has
been isolated \cite{Novoselov2004Science} and studied
\cite{Geim2005Nature,Kim2005Nature} (see
Refs.~\onlinecite{rev1,rev2} for a review). The two-dimensional (2D)
graphene honeycomb lattice has two atoms per unit cell. Its tight
binding band structure consists of the conduction and valence $\pi$
bands. These two bands touch each other and cross the Fermi level,
corresponding to  zero chemical potential $\mu$, in six $\mathbf{K}$
points located at the corners of the hexagonal 2D Brillouin zone,
but only two of them are inequivalent. The extended rhombic
Brillouin zone can be chosen to contain only these two points inside
the zone. In a field effect graphene device
\cite{Novoselov2004Science,Geim2005Nature,Kim2005Nature} electrons
can be introduced in the empty conduction band or holes in the
filled valence band through the application of a gate voltage and
thus the normally zero value of the chemical potential $\mu$ can be
changed continuously. In this paper we wish to consider the optical
sum rules described above for the specific case of graphene. Two
essential modifications arise. The existence of the two bands means
that interband transitions need to be accounted for in addition to
the usual intraband transitions of the previous discussion. Second,
the electronic dispersion curves near $\mathbf{K}$ points are linear
in momentum and the low-energy quasiparticle excitations are
described by the ``relativistic'' $(2+1)$ dimensional Dirac theory
\cite{Wallace1947PRev,Semenoff1984PRL,DiVincenzo1984PRB,Gonzales1993NP}
rather than the Schr\"{o}dinger equation.

The effective low-energy Dirac description of graphene turns out to
be insufficient for the derivation of the sum rules. This can be
easily understood from two examples. The RHS of
Eq.~(\ref{sum-rule-1band}) is normally related
(Refs.~\onlinecite{Millis:book,Marel:book,Benfatto:review,Carbotte:review})
to the diamagnetic term that is defined as the second derivative of
the Hamiltonian [see Eq.~(\ref{diamagnetic-term}) below] with
respect to the vector potential. This term is zero within the Dirac
approximation. On the other hand, it follows from the same Dirac
approximation that, in the high-frequency limit, the interband
contribution  to conductivity is  constant
\cite{Gusynin2006micro,Gusynin2007JPCM} (see also
Ref.~\onlinecite{Falkovsky2006})
\begin{equation}
\label{diagonal-high-frequency} \mbox{Re} \sigma_{xx}(\Omega) \simeq
\frac{\pi e^2}{2 h}, \qquad \Omega \gg \mu,T.
\end{equation}
The frequency $\Omega$ is unbound in the Dirac approximation,
although physically $\Omega$ should be well below the band edge.
This example indicates that when considering sum rules one should go
beyond the Dirac approximation.

The second example is that the cyclotron frequency $\omega_c$ for
the Dirac quasiparticles \cite{Zheng2002PRB} is defined in a
different way, $\omega_c = eB v^2_F/( c |\mu|)$, where $v_F$ is the
Fermi velocity. This definition follows from the fact that a
fictitious "relativistic" mass
\cite{Geim2005Nature,Kim2005Nature,Gusynin2006PRB} $m_c =
|\mu|/v_F^2$ plays the role of the cyclotron mass in the temperature
factor of the Lifshits-Kosevich formula for graphene
\cite{Sharapov2004PRB,Gusynin2005PRB}. This cyclotron frequency
diverges as $\mu \to 0$, also posing the question as what one should
use as a Hall frequency on the RHS of Eq.~(\ref{sum-rule-Hall}).

It turns out that these problems are resolved when a tight-binding
model is considered, but still due to the specifics of graphene, the
$f$-sum rule cannot be guessed from Eq.~(\ref{sum-rule-1band})
applied for the two-band case. Further important modifications to
the $f$-sum and Hall angle sum rule for graphene are expected and
found in this paper.

The paper is organized as follows. In Sec.~\ref{sec:model} we
introduce the tight-binding model and the necessary formalism. In
Secs.~\ref{sec:sum-diag} and \ref{sec:sum-Hall} we describe the
derivation of the sum rules for the diagonal and Hall conductivities
and consider the specific case of graphene. The partial sum rule for
the Hall angle is investigated in Sec.~\ref{sec:partial}. In
Sec.~\ref{sec:concl}, the main results of the paper are summarized.
Mathematical details are given in two Appendices.

\section{Tight-binding model and general representation for electrical conductivity}
\label{sec:model}

\subsection{Tight-binding model: Paramagnetic and diamagnetic parts}

The honeycomb lattice can be described in terms of two triangular
sublattices, A and B. The unit vectors of the underlying triangular
sublattice are chosen to be
\begin{equation}
\begin{split}
& \mathbf{a}_1= (a \sqrt{3}/2, a/2),\\
& \mathbf{a}_2= (a\sqrt{3}/2,-a/2),
\end{split}
\end{equation}
where the lattice constant $a = |\mathbf{a}_1| = |\mathbf{a}_2| =
\sqrt{3} a_{CC}$ and $a_{CC}$ is the distance between two nearest
carbon atoms. Any $A$ atom at the position $\mathbf{n}= \mathbf{a}_1
n_1 + \mathbf{a}_2 n_2$, where $n_1,n_2$ are integers, is connected
to its nearest neighbors on $B$ sites by the three vectors
$\pmb{\delta}_i$:
\begin{equation}
\begin{split}
& \pmb{\delta}_1 =-(\mathbf{a}_1+\mathbf{a}_2)/3, \\
& \pmb{\delta}_2 = 2/3 \mathbf{a}_1 - 1/3\mathbf{a}_2,\\
& \pmb{\delta}_3 =-1/3\mathbf{a}_1 + 2/3\mathbf{a}_2.
\end{split}
\end{equation}
We start with the simplest tight-binding description for $\pi$
orbitals of carbon in terms of the Hamiltonian
\begin{equation}
\label{Hamilton-lattice} H =
t\sum_{\mathbf{n},\pmb{\delta},\sigma}\left[a_{\mathbf{n},\sigma}^\dagger
 \exp \left( \frac{ie}{\hbar c}\pmb{\delta}\mathbf{A} \right)
b_{\mathbf{n}+\pmb{\delta},\sigma} + \mbox{c.c.}\right],
\end{equation}
where $t$ is the hopping parameter, $a_{\mathbf{n},\sigma}$ and
$b_{\mathbf{n}+ \mathbf{\delta},\sigma}$ are the Fermi operators of
electrons with spin $\sigma$ on $A$ and $B$ sublattices,
respectively. Since we are interested in the current response, the
vector potential $\mathbf{A}$ is introduced in the Hamiltonian
(\ref{Hamilton-lattice}) by means of the Peierls substitution
$a_{\mathbf{n},\sigma}^\dagger b_{\mathbf{m},\sigma}\rightarrow
a_{\mathbf{n},\sigma}^\dagger\exp\left(- \frac{i e}{\hbar
c}\int_{\mathbf{m}}^{\mathbf{n}}{\bf A}d
\mathbf{r}\right)b_{\mathbf{m},\sigma}$, which introduces the phase
factor $\exp (\frac{ie}{\hbar c}\pmb{\delta}\mathbf{A})$ in the
hopping term (see Ref.~\onlinecite{Millis:book} for a review). We
keep  the Planck constant $\hbar$ and the velocity of light $c$, but
set $k_B=1$.

Expanding the Hamiltonian (\ref{Hamilton-lattice}) to the second
order in the vector potential, one has
\begin{equation}
\label{Hamitonian-expand}
H=H_0-\sum_{\mathbf{n}}\left[\frac{1}{c}\mathbf{A}(\mathbf{n})
\mathbf{j}(\mathbf{n})-\frac{1}{2c^2}A_\alpha(\mathbf{n})\tau_{\alpha
\beta}(\mathbf{n})A_\beta(\mathbf{n})\right], \qquad \alpha,
\beta=1,2.
\end{equation}
The total current density operator is obtained by differentiating
the last equation with respect to $A_\alpha(\mathbf{n})$,
\begin{equation}
j_\alpha(\mathbf{n})=-\frac{\partial
H}{\partial\left(A_\alpha/c\right)}=j_\alpha^P(\mathbf{n})-
\tau_{\alpha \beta}(\mathbf{n})A_\beta(\mathbf{n})/c,
\end{equation}
and consists of the usual paramagnetic  part
\begin{equation}
\label{current-param} j_\alpha^P(\mathbf{n})=-\frac{ite}{\hbar}
\sum_{\pmb{\delta},\sigma}\delta_\alpha\left[a_{\mathbf{n},\sigma}^\dagger
b_{\mathbf{n}+\pmb{\delta},\sigma} -
b_{\mathbf{n}+\pmb{\delta},\sigma}^\dagger
a_{\mathbf{n},\sigma}\right]
\end{equation}
and diamagnetic part
\begin{equation}
\label{diamagnetic-term} \tau_{\alpha
\beta}(\mathbf{n})=\frac{\partial^2
H}{\partial\left(A_\alpha/c\right)\partial\left(A_\beta/c\right)}
=-\frac{te^2}{\hbar^2}
\sum_{\pmb{\delta},\sigma}\delta_\alpha\delta_\beta\left[a_{\mathbf{n},\sigma}^\dagger
b_{\mathbf{n}+\pmb{\delta},\sigma}
+b_{\mathbf{n}+\pmb{\delta},\sigma}^\dagger
a_{\mathbf{n},\sigma}\right].
\end{equation}

\subsection{Noninteracting Hamiltonian}

The noninteracting Hamiltonian $H_0$ in
Eq.~(\ref{Hamitonian-expand}) written in the momentum representation
reads
\begin{equation}
\label{H_0} H_0 =  \sum_{\sigma} \int_{BZ} \frac{d^2 k}{(2 \pi)^2}
\Upsilon_{\sigma}^\dagger (\mathbf{k}) \mathcal{H}_0
\Upsilon_{\sigma} (\mathbf{k}), \qquad \mathcal{H}_0 = \left(
  \begin{array}{cc}
    0 &  \epsilon(\mathbf{k}) e^{i \varphi (\mathbf{k})} \\
     \epsilon(\mathbf{k}) e^{-i \varphi (\mathbf{k})} & 0 \\
  \end{array}
\right),
\end{equation}
with $ \phi(\mathbf{k}) = t \sum_{\pmb{\delta}} e^{i \mathbf{k}
\pmb{\delta}} \equiv \epsilon(\mathbf{k}) e^{i \varphi
(\mathbf{k})}$. The dispersion law $\epsilon(\mathbf{k})$ is
\begin{equation}
\label{dispersion} \epsilon(\mathbf{k})=  t \sqrt{1+ 4\cos
\frac{\sqrt{3} k_x a}{2} \cos \frac{k_y a}{2} + 4 \cos^2 \frac{k_y
a}{2}}.
\end{equation}
In Eq.~(\ref{H_0}) we introduced the spinors
\begin{equation}
\label{spinor} \Upsilon_{\sigma} (\mathbf{k})= \left(
                                                 \begin{array}{c}
                                                   a_{\sigma} (\mathbf{k}) \\
                                                   b_{\sigma} (\mathbf{k})\\
                                                 \end{array}
                                               \right),
\end{equation}
with the operator $\Upsilon_{\sigma} (\mathbf{k})$ being the Fourier
transform of the spinor $\Upsilon_{\sigma}(\mathbf{n}) = \left(
                                          \begin{array}{c}
                                            a_{\mathbf{n},\sigma} \\
                                            b_{\mathbf{n},\sigma} \\
                                          \end{array}
                                        \right)$:
\begin{equation}
\label{Fourier} \Upsilon_{\sigma}(\mathbf{n}) =\sqrt{S}
\int_{BZ}\frac{d^2{\mathbf{k}}}{(2\pi)^2}\, e^{i \mathbf{k}
\mathbf{n}}  \Upsilon_{\sigma} (\mathbf{k}).
\end{equation}
Here  $S = \sqrt{3} a^2/2$ is the area of a unit cell and the
integration in Eqs.~(\ref{H_0}) and (\ref{Fourier}) goes over the
extended rhombic Brillouin zone (BZ) which is characterized by the
reciprocal lattice vectors $\mathbf{b}_1= 2\pi/a(1/\sqrt{3},1)$ and
$\mathbf{b}_2= 2\pi/a(1/\sqrt{3},-1)$.

We also add the term
\begin{equation}
\label{H-mu} H_0 \to H_0 - \mu \sum_\sigma \int_{BZ} \frac{d^2 k}{(2
\pi)^2} \Upsilon_{\sigma}^\dagger (\mathbf{k}) \hat{I}
\Upsilon_{\sigma}(\mathbf{k})
\end{equation}
with the chemical potential $\mu$ to the Hamiltonian $H_0$, so that
our subsequent consideration is based on the grand canonical
ensemble. The corresponding imaginary time Green's function (GF) is
defined as a thermal average
\begin{equation}
\label{GF-def} G_\sigma(\tau_1-\tau_2,\mathbf{n}_1 - \mathbf{n}_2) =
- \langle T_\tau \Upsilon_\sigma(\tau_1,\mathbf{n}_1)
\Upsilon_\sigma^{\dagger}(\tau_2,\mathbf{n}_2) \rangle
\end{equation}
and  its Fourier transform is
\begin{equation}
\label{GF-Fourier} G_\sigma(\tau_1-\tau_2,\mathbf{n}_1 -
\mathbf{n}_2) = T \sum_n \int_{BZ} \frac{d^2{\mathbf{k}}}{(2\pi)^2}
G(i \omega_n, \mathbf{k}) \exp[- i \omega_n (\tau_1 - \tau_2)+ i
\mathbf{k}(\mathbf{n}_1 - \mathbf{n}_2)],
\end{equation}
with
\begin{equation}
\label{GF-basic} G(i \omega_n,\mathbf{k}) = \frac{(i \omega_n + \mu)
\hat{I} + \sigma_+ \phi(\mathbf{k}) + \sigma_-
\phi^\ast(\mathbf{k})}{(i \omega_n + \mu)^2 -
\epsilon^2(\mathbf{k})},
 \qquad \omega_n = \pi (2n+1)T,
\end{equation}
where the matrix $\sigma_{\pm} = (\sigma_1 \pm i \sigma_2)/2$ made
from Pauli matrices operates in the sublattice space. In
Eq.~(\ref{GF-basic}) the spin label $\sigma$ is omitted, because in
what follows we neglect the Zeeman splitting and include a factor of
$2$ when necessary. The GF (\ref{GF-basic}) describes the
electronlike and holelike excitations with  energies
$E_{\pm}(\mathbf{k}) = \pm \epsilon(\mathbf{k})-\mu$, respectively.
The dispersion $\epsilon(\mathbf{k})$ near $\mathbf{K}$ points is
linear, $E_{\pm}({\bf p})=\pm \hbar v_F\sqrt{p_1^2+p_2^2} -\mu$,
where the wave vector ${\bf p}=(p_1,p_2)$ is now measured from the
$\mathbf{K}$ points and the Fermi velocity is $v_F = \sqrt{3}
ta/(2\hbar)$. Its experimental value
\cite{Geim2005Nature,Kim2005Nature} is $v_F  \approx 10^6
\mbox{m/s}$.

\subsection{Electrical conductivity}

The frequency-dependent electrical conductivity tensor
$\sigma_{\alpha \beta}(\Omega)$ is calculated using the Kubo formula
\cite{Millis:book,Marel:book}
\begin{equation}\label{Kubo}
\sigma_{\alpha \beta}(\Omega)= \frac{K_{\alpha
\beta}(\Omega+i0)}{-i(\Omega+i0)}, \qquad K_{\alpha
\beta}(\Omega+i0) \equiv \frac{\langle\tau_{\alpha
\beta}\rangle}{V}+ \frac{\Pi_{\alpha \beta}^R(\Omega+i0)}{\hbar V},
\end{equation}
where the retarded correlation function for currents is given by
\begin{equation}
\Pi_{\alpha \beta}^R(\Omega)=\int_{-\infty}^\infty dt\,e^{i\Omega
t}\Pi_{\alpha \beta}^R(t),\quad \Pi_{\alpha
\beta}^R(t)=-i\theta(t){\rm
Tr}\left(\hat{\rho}[J_\alpha(t),J_\beta(0)]\right),
\end{equation}
$V$ is the volume of the system, $\hat{\rho}=\exp(-\beta H_0)/Z$ is
the density matrix of the grand canonical ensemble, $\beta=1/T$ is
the inverse temperature, $Z={\rm Tr}\exp(-\beta H_0)$ is the
partition function, and $J_\alpha$ are the total paramagnetic
current operators with
\begin{equation}
\label{current_t=0}
J_\alpha(t)=e^{iHt/\hbar}J_\alpha(0)e^{-iHt/\hbar}, \qquad
J_\alpha(t)=\sum_{\mathbf{n}} j_\alpha^P(t,\mathbf{n}),
\end{equation}
expressed via the paramagnetic current density
(\ref{current-param}). Using the representation for $\Pi_{\alpha
\beta}(\Omega)$ in terms of the matrix elements of the current
operator $J_\alpha(t=0)$, one can find \cite{Shastry1993PRL} the
high-frequency, $\Omega \to \infty$ asymptotic of the Hall
conductivity,
\begin{equation}
\label{sigmaxy-high_freq}
\sigma_{xy}(\Omega)=\frac{i}{V\hbar\Omega^2}\left\{\langle[J_x,J_y]\rangle+\frac{1}{(\hbar\Omega)^2}
\langle[[[J_x,H],H]J_y]\rangle+O\left(\frac{1}{(\hbar\Omega)^4}\right)\right\},
\end{equation}
while the longitudinal conductivity is
\begin{equation}
\label{sigmaxx-high_freq}
\sigma_{xx}=\frac{1}{iV\Omega}\left\{-\langle\tau_{xx}\rangle+\frac{1}{(\hbar\Omega)^2}
\langle[[J_x,H],J_x]\rangle+
O\left(\frac{1}{(\hbar\Omega)^4}\right)\right\}.
\end{equation}
Here $[,]$ is the commutator. Below in Secs.~\ref{sec:sum-diag} and
\ref{sec:sum-Hall} we will use Eqs.~(\ref{Kubo}),
(\ref{sigmaxy-high_freq}), and (\ref{sigmaxx-high_freq}) to outline
the formal derivation of the sum rules.

\section{Diagonal optical conductivity sum rule}
\label{sec:sum-diag}

The optical conductivity sum rule is a consequence of gauge
invariance and causality. Gauge invariance dictates the way that the
vector potential enters Eq.~(\ref{Hamilton-lattice}) and,
respectively, determines the diamagnetic and paramagnetic terms in
the expansion (\ref{Hamitonian-expand}) as well as the form of Kubo
formula (\ref{Kubo}). The causality implies that the conductivity,
Eq.~(\ref{Kubo}) satisfies the Kramers-Kr\"onig (KK) relation
\begin{equation} \label{KK-relation_general}
\sigma_{\alpha \beta}(\Omega)=\frac{1}{\pi i} \mathrm{P}
\int_{-\infty}^\infty \frac{d\omega\,\sigma_{\alpha
\beta}(\omega)}{\omega-\Omega}.
\end{equation}
Combining together the high-frequency limit of the KK relation
(\ref{KK-relation_general}),
\begin{equation}
\mathrm{Im} \sigma_{\alpha \beta}(\Omega)= \frac{1}{\pi \Omega}
\mathrm{P}\int_{-\infty}^\infty d\omega \mathrm{ Re}\sigma_{\alpha
\beta}(\omega), \qquad \Omega \to \infty,
\end{equation}
and the asymptotic (\ref{sigmaxx-high_freq}), we arrive at the sum
rule
\begin{equation}
\label{sum-rule-sigma} \frac{1}{\pi}\int\limits_{-\infty}^\infty
d\Omega \mathrm{ Re} \sigma_{xx}(\Omega)=
\frac{\langle\tau_{xx}\rangle}{V}.
\end{equation}
Taking into account that $\mathrm{Re}\sigma_{xx}(\Omega)$ is an even
function of $\Omega$, we observe that for a single tight-binding
band $\langle\tau_{\alpha \beta}\rangle/V$ corresponds to the RHS of
Eq.~(\ref{sum-rule-1band}). Also defining the plasma frequency
$\omega_P$ via $\omega_P^2/(4\pi)\delta_{\alpha \beta} \equiv
\langle\tau_{\alpha \beta}\rangle/V$ we can rewrite the sum rule
(\ref{sum-rule-sigma}) in the form (\ref{sum-rule-free}). Below we
calculate this term for the Hamiltonian (\ref{Hamilton-lattice})
with a nearest neighbor hopping on the hexagonal graphene lattice.

\subsection{Explicit form of the diamagnetic term}
\label{sec:diamagn}

The diamagnetic or stress tensor $\langle \tau_{\alpha
\beta}\rangle$ in the Kubo formula (\ref{Kubo}) is a thermal average
of Eq.~(\ref{diamagnetic-term})
\begin{equation}
\langle \tau_{\alpha \beta}\rangle =  \langle \sum_{\mathbf{n}}
\tau_{\alpha \beta} (\mathbf{n})  \rangle.
\end{equation}
This term is calculated in Appendix~\ref{sec:A} and is given by
\begin{equation} \label{sum-rule-phase}
\frac{\langle\tau_{\alpha
\alpha}\rangle}{V}=\frac{2e^2}{\hbar^2}\int_{BZ}\frac{d^2{\bf
k}}{(2\pi)^2}\left[n_F\left( \epsilon(\mathbf{k})
\right)-n_F(-\epsilon(\mathbf{k}))\right]\left[\frac{\partial^2}{\partial
k_\alpha^2} -\left(\frac{\partial\varphi(\mathbf{k})}{\partial
k_\alpha}\right)^2\right]\epsilon(\mathbf{k}),
\end{equation}
where $n_{F}(\omega) =1/[\exp((\omega-\mu)/T)+1]$ is the Fermi
distribution. Because the graphene structure contains two atoms per
unit cell (two sublattices), there are two bands in the BZ which
correspond to positive- and negative-energy Dirac cones. The
momentum integration in Eq.~(\ref{sum-rule-phase}) is over the
entire BZ and the thermal factors $n_F(\epsilon(\mathbf{k}))$ and
$n_F(-\epsilon(\mathbf{k}))$ refer to the upper and lower Dirac
cones, respectively. We note that a simple generalization of
Eq.~(\ref{sum-rule-1band}) for a two band case would miss the term
with the derivative of the phase,
$\left(\partial\varphi(\mathbf{k})/\partial k_\alpha\right)^2$. This
term occurred due to the fact that the Peierls substitution was made
in the initial Hamiltonians (\ref{Hamilton-lattice}) and (\ref{H_0})
rather than after the diagonalization of Eq.~(\ref{H_0}). The second
comment on Eq.~(\ref{sum-rule-phase}) [see also
Eq.~(\ref{tau-average}) in Appendix~\ref{sec:A}] is that
$\langle\tau_{\alpha \alpha}\rangle$ vanishes if
$\epsilon(\mathbf{k})$ is taken in the linear approximation. This
reflects the absence of the diamagnetic term in the Dirac
approximation. The correct way is firstly to take the derivatives in
Eq.~(\ref{sum-rule-phase}). This is done in Eq.~(\ref{derivative})
in Appendix~\ref{sec:A} and leads to the final result
\begin{equation} \label{sum-rule-main}
\frac{\langle\tau_{\alpha \alpha}\rangle}{V}
=-\frac{e^2a^2}{3\hbar^2}\int_{BZ}\frac{d^2{\bf
k}}{(2\pi)^2}\left[n_F(\epsilon(\mathbf{k})) -
n_F(-\epsilon(\mathbf{k}))\right] \epsilon(\mathbf{k}).
\end{equation}
Eq.~(\ref{sum-rule-main}) is equivalent to
Eq.~(\ref{sum-rule-phase}). Note that $\langle\tau_{\alpha
\alpha}\rangle$ is always positive and does not depend on the
arbitrary choice of the sign before $t$ in
Eq.~(\ref{Hamilton-lattice}). Now Eq.~(\ref{sum-rule-main}) is
$e^2/\hbar^2$ times $-2/(3\sqrt{3})(\sim - 0.39)$ of the kinetic
energy per atom instead of $-1/2$ for the usual square lattice. At
zero temperature for $\mu>0$ (to be specific) the lower Dirac cone
is full and, in the conductivity, only interband transitions are
possible for these electrons. Similarly, the electrons in the upper
Dirac cone can undergo only intraband transitions. In the above
sense the first thermal factor in Eq.~(\ref{sum-rule-main})
corresponds to intraband and the second to interband at $T=0$.

It is useful to separate explicitly the contribution
$\langle\tau_{xx}(\mu=T=0)\rangle$ of the Dirac sea
 from Eq.~(\ref{sum-rule-main}):
\begin{equation}
\langle\tau_{xx}\rangle = \langle\tau_{xx}(\mu=T=0)\rangle +
\langle\tau_{xx}^{eh}(\mu,T)\rangle,
\end{equation}
where
\begin{equation}
\label{Dirac-sea} \frac{\langle\tau_{xx}(\mu=T=0)\rangle}{V}
=-\frac{e^2a^2}{3\hbar^2}\int_{BZ}\frac{d^2{\bf k}}{(2\pi)^2} (-
\epsilon(\mathbf{k}))
\end{equation}
is the contribution of the Dirac sea (the energy of the filled
valence band) and
\begin{equation}
\label{e-h}
\frac{\langle\tau_{xx}^{eh}(\mu,T)\rangle}{V}=-\frac{e^2a^2}{3\hbar^2}\int_{BZ}\frac{d^2{\bf
k}}{(2\pi)^2}\left[n_F(\epsilon(\mathbf{k})) + 1-
n_F(-\epsilon(\mathbf{k}))\right] \epsilon(\mathbf{k})
\end{equation}
is the electron-hole contribution. Expressions
(\ref{sum-rule-main}), (\ref{Dirac-sea}), and (\ref{e-h}) contain
the energy $\epsilon(\mathbf{k})$ as happens also for the case of
the square lattice with nearest neighbor hopping mentioned in the
Introduction.

The numerical calculation of the Dirac sea contribution
(\ref{Dirac-sea}) with the full dispersion (\ref{dispersion}) gives
\begin{equation}
\label{Dirac-sea-num} \frac{\langle\tau_{xx}(\mu=T=0)\rangle}{V} =
\alpha \frac{e^2 t}{\hbar^2}, \qquad \alpha \approx 0.61.
\end{equation}
The same answer also follows from the linearized Dirac approximation
with the trigonal density of states if the band width $W$ is taken
to be $W= \sqrt{\sqrt{3} \pi}t$. The electron-hole contribution
(\ref{e-h}) can be estimated analytically in the linear
approximation for the dispersion law,
\begin{equation}
\label{e-h-Li} \frac{\langle\tau_{xx}^{eh}(\mu,T)\rangle}{V}=
\frac{2e^2a^2 T^3}{3\pi\hbar^4v_F^2} \left[{\rm
Li}_3(-e^{\mu/T})+{\rm Li}_3(-e^{-\mu/T})\right],
\end{equation}
where $\mathrm{Li}_{3}(z)$ is the polylogarithmic function
\cite{Levin.book}. This shows that for $\mu=0$ the temperature
dependence of the diagonal conductivity sum rule is $\sim T^3$, in
contrast to a well-known $T^2$ dependence
\cite{Millis:book,Marel:book,Benfatto:review,Carbotte:review}. We
note, however, that because $\langle\tau_{xx}^{eh}(\mu=0,T)\rangle/V
\sim -T^3/t^2$ which is small, the $T^3$ behavior is unlikely to be
observed. On the other hand, using asymptotics of the
polylogarithmic function for $|\mu| \gg T$, Eq.~(\ref{e-h-Li}) can
be written in the form
\begin{equation}
\label{e-h-approx} \frac{\langle\tau_{xx}^{eh}(\mu,T)\rangle}{V}=-
\frac{e^2a^2}{9\pi\hbar^4v_F^2} \left[|\mu|^3+\pi^2|\mu|T^2 \right].
\end{equation}
The $|\mu|^3$ behavior at $T=0$ is easily understood from
Eq.~(\ref{e-h}) in which case it reduces to an integral over energy
ranging from $0$ to $\mu$ of the density of states (DOS) which is
proportional to $|\epsilon|$ in graphene. The integrand is therefore
proportional to $\epsilon^2$ leading directly to the $|\mu|^3$
dependence. This cubic power law reflects directly the Dirac nature
of the electronic dispersion relation encoded in the linear in the
$\epsilon$ DOS. At finite $T$ the DOS in Eq.~(\ref{e-h})  for
$\mu=0$ provides an additional factor of $T$ as compared with the
conventional case, leading to $T^3$ rather than $T^2$ behavior. On
the other hand, for $|\mu| \gg T$ the DOS can be evaluated at $\mu$,
leading to a $|\mu| T^2$ law. All these results follow directly from
a linear dispersion of the massless Dirac quasiparticles.

We note that the $|\mu|^3$ behavior is more likely to be observed
than the $T^3$ dependence by varying the gate voltage $V_g$. Using
Eq.~(\ref{rho}) for carrier imbalance $\rho$ which is proportional
to $V_g$, we obtain that $\langle\tau_{xx}^{eh}(V_g)\rangle/V \sim
-|V_g|^{3/2}$. Note that as one can see from Eq.~(\ref{e-h-approx}),
the electron-hole contribution $\langle\tau_{xx}^{eh}(\mu,T)\rangle$
in Eq.~(\ref{e-h}) is negative and it has to be added to
Eq.~(\ref{Dirac-sea}) to get the total non-negative contribution to
the optical sum. If at $T=0$ we take the chemical potential to fall
at the top of the band, so that the positive energy Dirac cone is
completely occupied, Eq.~(\ref{e-h}) reduces to
Eq.~(\ref{Dirac-sea}) except for a difference in sign. The two
contributions cancel, because a full band cannot absorb. The same
result follows from Eq.~(\ref{e-h-approx}) when $\mu$ is taken to be
the energy cutoff on the band-width $W= \sqrt{\sqrt{3} \pi}t$.  We
also note from Eq.~(\ref{e-h-approx}) that for $|\mu| \gg T$ we
recover the already mentioned $T^2$ law
\cite{Millis:book,Marel:book,Benfatto:review,Carbotte:review}.

\section{Hall angle sum rule}
\label{sec:sum-Hall}

We now consider the optical Hall angle (\ref{t_H}) that plays the
role of the response function to an injected current rather than an
applied field:
\begin{equation}\label{response-Hall}
j_x(\Omega) = \sigma_{xy}(\Omega) E_y(\Omega) =
t_H(\Omega)j_y(\Omega), \qquad j_y(\Omega) = \sigma_{xx}(\Omega)
E_y(\Omega).
\end{equation}
It was proved in Ref.~\onlinecite{Drew1997PRL} that the response
function $t_H(\Omega)$ satisfies the KK relation
\begin{equation}\label{t_H-KK}
t_H(\Omega) = \frac{1}{\pi i} \mbox{P}\int_{-\infty}^{\infty} d
\omega \frac{1}{\omega - \Omega}t_H(\omega).
\end{equation}
Multiplying Eq.~(\ref{t_H-KK}) by $- i \Omega$ and taking the limit
$\Omega \to \infty$, we obtain the sum rule
\begin{equation}\label{t_H-rule}
\frac{1}{\pi}\int_{-\infty}^{\infty} d \omega t_H(\omega) =
\omega_H,
\end{equation}
with the Hall frequency
\begin{equation}
\label{omega_H} \omega_H \equiv \lim_{|\Omega|\to \infty}[- i \Omega
t_H(\Omega)].
\end{equation}
The Hall angle sum rule (\ref{sum-rule-Hall}) follows from
Eq.~(\ref{t_H-rule}) after we take into account that the real and
imaginary parts of $t_H(\omega)$ are even and odd functions of
$\Omega$, respectively. Microscopic considerations based on the Kubo
formula (\ref{Kubo}) show that the high-frequency limit
(\ref{omega_H}) exists and is given by
\begin{equation}
\label{omega_H-micro} i \omega_H = \lim_{|\Omega|\to \infty}
\frac{\Omega K_{xy}(\Omega)}{K_{xx}(\Omega)} = \frac{\langle
[J_x(t=0), J_y(t=0)]\rangle}{\hbar \langle \tau_{xx} \rangle},
\end{equation}
where in the last equality we used the high-frequency asymptotics,
Eqs.~(\ref{sigmaxy-high_freq}) and (\ref{sigmaxx-high_freq}).

The commutator $I_{x,y} = \langle [J_x(t=0), J_y(t=0)]\rangle$ is
calculated in Appendix~\ref{sec:B}, where we obtain
\begin{equation}
\label{commutator-final} I_{x,y}  = - i \frac{e^2 a^4 t^2 \hbar e
B}{4 \hbar^4 c} \epsilon_{xy} V \rho .
\end{equation}
Here $\rho$ is the carrier imbalance ($\rho = n_e - n_h$, where
$n_e$ and $n_h$ are the densities of electrons and holes,
respectively), and $\epsilon_{ab}$ is antisymmetric tensor. The
carrier imbalance for $B=T=0$ and in the absence of impurities is
\begin{equation} \label{rho} \rho = \frac{\mu^2 \mbox{sgn}
\mu}{\pi \hbar^2 v_F^2}.
\end{equation}
Substituting Eqs.~(\ref{commutator-final}) and (\ref{Dirac-sea-num})
in (\ref{omega_H-micro}) we finally  obtain
\begin{equation}
\label{omega_H-final} \omega_H = - \frac{1}{4\alpha} \frac{eB}{c}
\frac{t a^2}{ \hbar^2}\rho a^2.
\end{equation}
Since $t a^2/\hbar^2$ has the dimensionality of the inverse mass and
$\rho a^2$ is dimensionless, Eq.~(\ref{omega_H-final}) has the
correct dimensionality of the cyclotron frequency. Substituting
Eq.~(\ref{rho}) and expressing $v_F$ via $t$, one can rewrite
\begin{equation}
\label{omega_H-final-K} \omega_H = - \frac{4\,\mbox{sgn}\,(eB)}{9
\pi\alpha} L^2(B) \frac{\mu^2 \mbox{sgn} \mu}{\hbar t^3 }.
\end{equation}
Here $L(B)= \sqrt{|eB| \hbar v_F^2/c}$ is the Landau scale which  in
temperature units is equal to $L^2(B)[\mbox{K}^2] = 8.85 \times
10^{-8} \mbox{K}^2 v_F^2(\mbox{m/s}) B(T) $ and $\hbar = 7.638
\times 10^{-12} K \cdot s$.

As mentioned in the Introduction, in the recent interpretation of
Shubnikov de Haas measurements  a gate voltage-dependent cyclotron
mass was introduced \cite{Geim2005Nature,Kim2005Nature} through the
relationship $|\mu| = m_c v_F^2$. If this is used in
Eq.~(\ref{omega_H-final-K}) we get
\begin{equation}
\label{omega_H-filled} \omega_H = -\frac{eB}{c m_c}
\left(\frac{\mu}{1.62 t}\right)^3.
\end{equation}
Since a full upper Dirac band corresponds to a value $\mu =
W=\sqrt{\sqrt{3} \pi} t \simeq 2.33 t$ (see the end of
Sec.~\ref{sec:diamagn}), in this case formula (\ref{omega_H-filled})
resembles the formula $\omega_H = \omega_c = eB/mc$ from
Ref.~\onlinecite{Drew1997PRL} for two-dimensional electron gas with
$m_c$ replacing the free electron mass. In graphene, however, $m_c$
varies as the square root of the carrier imbalance $|\rho|$ and the
two cases look the same only formally.

Finally, we notice that when the spectrum becomes gapped with $E =
\sqrt{\hbar^2 v_F^2(p_1^2 + p_2^2) + \Delta^2}$, the carrier
imbalance is
\begin{equation}
\label{rho-gap} \rho = \frac{1}{\pi \hbar^2 v_F^2}(\mu^2 -
\Delta^2)\theta(\mu^2 - \Delta^2)\mbox{sgn} \mu
\end{equation}
This implies that the gap $\Delta$ can be extracted from the change
in $\omega_H$ obtained from  magneto-optical measurements. This kind
of measurement  which reveals gapped behavior has  already been done
on the underdoped high-temperature superconductor YBa$_2$Cu$_3$O$_{6
+ x}$ \cite{Rigal2004PRL}. The recent measurements done in epitaxial
graphite \cite{Sadowski2006} and in highly oriented pyrolytic
graphite \cite{Li2006PRB} lead us to expect that this experiment
should be possible for graphene. Below we restrict ourselves to the
$\Delta=0$ case.

\section{Partial spectral weight for the Hall angle sum}
\label{sec:partial}

So far we have considered only the  complete optical sum involving
integration of the optical spectral weight over the entire band. The
temperature dependence of such a sum has been central to many recent
studies \cite{Marel:book,Benfatto:review,Carbotte:review} focused on
the possibility of kinetic energy driven superconductivity in the
cuprates. The effects are small and the experiment is difficult. The
conductivity is needed up to high energies and there is no
unambiguous criterion to decide where the band of interest may end.
On the other hand the partial optical sum to a definite upper limit
$\Omega_m$ can give important information on the approach to the
complete optical sum. It has also proved very useful in
understanding the spectral weight redistribution with temperature or
phase transition (see, e.g., the recent works in
Refs.~\onlinecite{Hwang2006} and \onlinecite{Bontemps2006}). A
discussion of the effect of an external magnetic field on the
partial optical sum (or weight of the main peak) for the
longitudinal conductivity in epitaxial graphite can be found in
Ref.~\onlinecite{Sadowski2006}. While the optical Hall conductivity
is not as widely studied as the longitudinal conductivity, many new
studies (see, e.g., Ref.~\onlinecite{Rigal2004PRL}) have shown its
usefulness as it gives information on the change in microscopic
interactions brought about by $B$. The available experimental data
for graphite \cite{Li2006PRB}  already contain information about
optical Hall conductivity, and when the same measurements are done
on graphene, they can be directly compared with the results
discussed below.

We saw in our previous paper \cite{Gusynin2007JPCM} that the
longitudinal conductivity
\begin{equation}
\label{W_xx} W_{xx}(\Omega_m) = \int_{0}^{\Omega_m} d \Omega
\mbox{Re}\, \sigma_{xx}(\Omega)
\end{equation}
showed (see Fig.~7 in Ref.~\onlinecite{Gusynin2007JPCM}) plateaux
with the steps corresponding to the various peaks in the diagonal
conductivity. Here we are interested in the corresponding quantity
associated with the Hall angle $\mbox{Re}\, t_H(\Omega)$ and
consider the weight
\begin{equation}
\label{W_xy} W(\Omega_m) = \int_{0}^{\Omega_m} d \Omega \mbox{Re}\,
t_{H}(\Omega).
\end{equation}
It is instructive to begin with a discussion of the frequency
dependence of $\mbox{Re}\, t_H(\Omega)$. As already mentioned after
Eq.~(\ref{t_H-rule}), because the imaginary part of both
$\sigma_{xy} (\Omega)$ and $\sigma_{xx} (\Omega)$ is an odd function
of $\Omega$, only the real part of $t_H(\Omega)$ contributes to the
sum rule (\ref{sum-rule-Hall}). Thus we will need to consider only
$\mbox{Re}\, t_H(\Omega)$ defined by Eq.~(\ref{t_H}). General
expressions for the complex conductivities $\sigma_{xx} (\Omega)$
and $\sigma_{xy} (\Omega)$, derived in the Dirac approximation, are
given by Eqs.~(9), (11) and (10), (12), respectively, of our
previous paper \cite{Gusynin2007JPCM}.

In what follows we will consider explicitly possible experimental
configurations. For a fixed cutoff ($\Omega_m$) sweeping the
magnetic field gives direct information on the change in optical
spectral weight, in this frequency region, brought about by $B$. For
the field effect transistor configuration used in
Refs.~\onlinecite{Novoselov2004Science,Geim2005Nature,Kim2005Nature}
the chemical potential $\mu$ can be changed through an adjustment of
the gate voltage at fixed $B$ and $\Omega_m$. Another configuration
that might be considered is to fix $B$ and $\mu$ and change
$\Omega_m$. These three cases serve to illustrate what can be
expected in experiments.

\begin{figure}[h]
\centering{
\includegraphics[width=8cm]{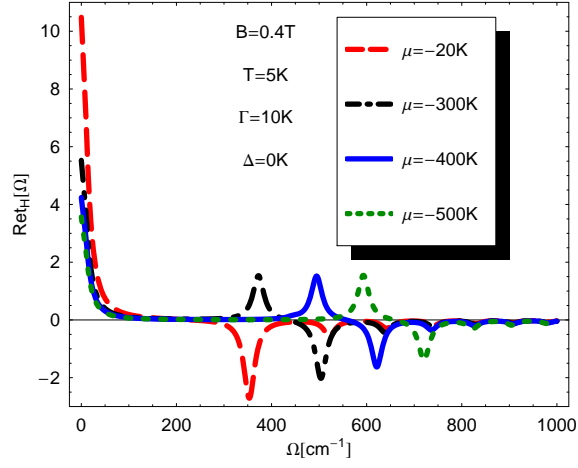}
} \caption{(Color online)  Real part of the Hall angle, $\mbox{Re}
t_H(\Omega)$ at $B=0.4 \, \mbox{T}$, temperature $T= 5 \mbox{K}$,
scattering rate $\Gamma =10 \, \mbox{K}$ for four values of chemical
potential. Long dashed (red) $\mu = -20 \, \mbox{K}$, dash-dotted
(black) $\mu = -300 \, \mbox{K}$, solid (blue) $\mu = -400 \,
\mbox{K}$, short dashed (green) $\mu = -500 \, \mbox{K}$. }
\label{fig:1}
\end{figure}

In Fig.~\ref{fig:1} we show results for the frequency dependence of
$\mbox{Re}\, t_H(\Omega)$ for a case with external magnetic field
$B=0.4\, \mbox{T}$, at temperature $T = 5 \, \mbox{K}$ and impurity
scattering rate $\Gamma = 10 \, \mbox{K}$. Accordingly, the energies
of the Landau levels are $M_n = \sqrt{2 n L^2(B)}$ [see also Eq.~(5)
in Ref.~\onlinecite{Gusynin2007JPCM}]--- viz., $M_1 \simeq 265\,
\mbox{K}$, $M_2 \simeq 375\, \mbox{K}$, $M_3 \simeq 460\, \mbox{K}$,
and $M_4 \simeq 531\, \mbox{K}$, respectively. Thus the long dashed
(red) curve with $\mu = -20 \, \mbox{K}$ is for $|\mu| < M_1$, the
dash-dotted (black) curve with $\mu = -300 \, \mbox{K}$ is for $M_1
< |\mu| < M_2$, the solid (blue) line with
 $\mu = -400 \, \mbox{K}$ is for $M_2 < |\mu| < M_3$ and
the short dashed (green) curve with $\mu = -500 \, \mbox{K}$ is for
$M_3 < |\mu| < M_4$. The values chosen for all the parameters used
in Fig.~\ref{fig:1} are quite reasonable. For a magnetic field $0.4
\mbox{T}$ the various Landau energies fall within the range of
available optical spectrometers and this is one value of $B$ used in
Ref.~\onlinecite{Sadowski2006} on the optics of several-layer
graphite. The value of the broadening parameter $\Gamma$ is not so
well known, but our choice is reasonable when compared with the line
broadening seen in experiments.\cite{Sadowski2006} Higher values
would simply broaden the lines seen in Fig.~\ref{fig:1}. Further in
field-effect graphene devices
\cite{Novoselov2004Science,Geim2005Nature,Kim2005Nature} the
chemical potential can be easily varied from $-3600\, \mbox{K}$ to
$3600\, \mbox{K}$ simply by changing the gate voltage. Low
temperatures can also be easily achieved. Starting with the long
dashed (red) curve of Fig.~\ref{fig:1} we notice a Drude-like
behavior of $\mbox{Re}\, t_H(\Omega)$ at small $\Omega$, followed by
a region where it is small. It then changes sign, after which it
exhibits a negative peak around $375 \, \mbox{cm}^{-1}$. This is
followed by a further series of small peaks, all negative, which
decay in amplitude as the frequency $\Omega$ increases. The
frequencies of the peaks do not correspond to the sum or differences
of two adjacent Landau level energies $M_n$ as do the peaks in both
$\mbox{Re} \, \sigma_{xx}(\Omega)$ and $\mbox{Im} \,
\sigma_{xy}(\Omega)$ seen in Ref.~\onlinecite{Gusynin2007JPCM}.
Rather they fall, approximately, at energies between two peaks in
$\mbox{Re}\, \sigma_{xx}(\Omega)$ (see Fig.~1 in
Ref.~\onlinecite{Gusynin2007JPCM} taking into account that it is
plotted for $B=1\, \mbox{T}$), where the $\mbox{Im}\,
\sigma_{xx}(\Omega)$ also crosses zero and the denominator in
$\mbox{Re}\,t_H = \mbox{Re}(\sigma_{xy}/\sigma_{xx})$ which is equal
to $(\mbox{Re}\, \sigma_{xx}(\Omega))^2 + (\mbox{Im}\,
\sigma_{xx}(\Omega))^2 $ consequently has a minimum. The numerator
$\mbox{Re} \, \sigma_{xx}(\Omega) \mbox{Re} \, \sigma_{xy}(\Omega) +
\mbox{Im}\, \sigma_{xx}(\Omega) \mbox{Im}\, \sigma_{xy}(\Omega)$
plays little role in the peak positions, but determines their sign
which can be positive or negative as seen in the other three curves
(for larger values of chemical potential). In this case the first
peak is positive and this is followed by, decaying in amplitude, a
sequence of higher-energy peaks, all of which are negative. Note
that the position of the first positive peak in each of the
dash-dotted, solid, and dashed curves moves progressively to higher
energy as the chemical potential crosses a new Landau level. This
happens because, as we have already stated, the peaks in
$\mbox{Re}\, t_H(\Omega)$ fall in the between two consecutive peaks
in $\mbox{Re} \, \sigma_{xx}(\Omega)$ (for $M_{N} < |\mu| <
M_{N+1}$, the first interband peak in $\mbox{Re} \,
\sigma_{xx}(\Omega)$ falls at $M_N + M_{N+1}$).

Returning to the small-$\Omega$ behavior of $\mbox{Re}\,
t_H(\Omega)$, where a Drude-like behavior was noted, we see that the
weight under this peak is progressively depleted as $\mu$ crosses
through larger Landau level energies. These features can be
understood qualitatively from approximate expressions for
$\mbox{Re}\, t_H(\Omega)$ obtained by keeping only the leading term
in the formulas for both the longitudinal $\sigma_{xx}(\Omega)$ and
Hall $\sigma_{xy}(\Omega)$ conductivity. In
Ref.~\onlinecite{Gusynin2007JPCM} we provided expressions (11) and
(12) which we will not repeat here. Both involve a sum over Landau
levels $n=0,1,2,\ldots$. If $|\mu|$ falls in the interval
$]M_0,M_1[$, retaining only the $n=0$ contribution to the sum,  we
get for $T=0$
\begin{equation}
\label{sigma_xx-approx} \sigma_{xx}(\Omega) = \frac{e^2 v_F^2 |eB|
}{\pi c i} \frac{2(\Omega + 2 i \Gamma)}{M_1} \frac{1}{M_1^2 -
(\Omega + 2 i \Gamma)^2}
\end{equation}
and
\begin{equation}
\label{sigma_xy-approx} \sigma_{xy}(\Omega) =- \frac{e^2 v_F^2 eB\,
\mbox{sgn}\, \mu }{\pi c}\frac{2}{M_1^2 - (\Omega + 2 i \Gamma)^2}.
\end{equation}
It follows from Eqs.~(\ref{sigma_xy-approx}) and
(\ref{sigma_xx-approx}) that
\begin{equation}
\label{t_H-Drude} t_H(\Omega) = - \frac{i \sqrt{2\hbar
|eB|v_F^2/c}\, \mbox{sgn}\, (eB)\, \mbox{sgn}\, \mu  }{\Omega+ 2 i
\Gamma}, \qquad \mbox{Re}\, t_H(\Omega) = - \frac{M_1 2 \Gamma\,
\mbox{sgn}\, (eB)\, \mbox{sgn}\, \mu }{\Omega^2 + 4 \Gamma^2}.
\end{equation}
This shows that while $\mbox{Re} \, \sigma_{xx}(\Omega)$ and
$\mbox{Im} \, \sigma_{xy}(\Omega)$ are peaked near $\Omega = M_1$,
the small $\Omega$ limit of $\mbox{Re}\, t_H(\Omega)$ is indeed
Drude in shape with width $2 \Gamma$. Also its height at $\Omega =0$
and $\mu<0$ is $M_1/2 \Gamma$ which is set by the inverse scattering
rate $2 \Gamma$ and by the Landau scale $L(B)$, because $M_1 =
\sqrt{2} L(B)$.

On the other hand, for $\mu \in ]M_{N}, M_{N+1}[$  the first nonzero
term in the expression for the Hall angle $t_H(\Omega)$ is for
$n=N$. Retaining only this contribution to the conductivities (11)
and (12) from Ref.~\onlinecite{Gusynin2007JPCM} and defining
\begin{equation}
A_{N}^{\pm} = \frac{1}{(M_{N+1} \pm M_{N})^2 - (\Omega + 2 i
\Gamma)^2}
\end{equation}
for $T=0$ we get
\begin{eqnarray}
\sigma_{xx}(\Omega)&=&\frac{e^{2}v_{F}^{2}|eB|(\Omega+2i\Gamma)}{\pi ci}
\left(\frac{A_{N}^{-}}{M_{N+1}-M_{N}}+\frac{A_{N}^{+}}{M_{N+1}+M_{N}}\right),\\
\sigma_{xy}(\Omega)&=&-\frac{e^{2}v_{F}^{2}eB}{\pi c}\left(A_{N}^{-}+A_{N}^{+}\right).
\end{eqnarray}
Hence for $t_{H}$ we obtain the Drude term
\begin{equation}
t_H(\Omega) = - i \frac{\mbox{sgn}\,(eB)\, \mbox{sgn}\, \mu }{\Omega
+ 2 i \Gamma} \frac{A_{N}^{-} + A_{N}^{+}}{A_{N}^{-}/(M_{N+1} -M_N)
+ A_{N}^{+}/(M_{N+1} + M_N)}.
\end{equation}
It follows directly from this formula that $\mbox{Re}\, t_H(\Omega)$
in the limit of large $N$ is given approximately by
\begin{equation}
\label{t_H-Drude-N} \mbox{Re}\, t_H(\Omega) \cong - \frac{1}{2
\sqrt{N}} \frac{M_1 2 \Gamma\, \mbox{sgn}\, (eB)\, \mbox{sgn}\,
\mu}{\Omega^2 + 4 \Gamma^2}, \qquad N \gg 1.
\end{equation}
The value $\mbox{Re}\, t_H(\Omega=0) = M_1/(4 \sqrt{N} \Gamma)$ is
roughly verified in our numerical work, where larger $N$ corresponds
to larger chemical potential, and these results serve as a guide to
our numerical work.

\begin{figure}[h]
\centering{
\includegraphics[width=8cm]{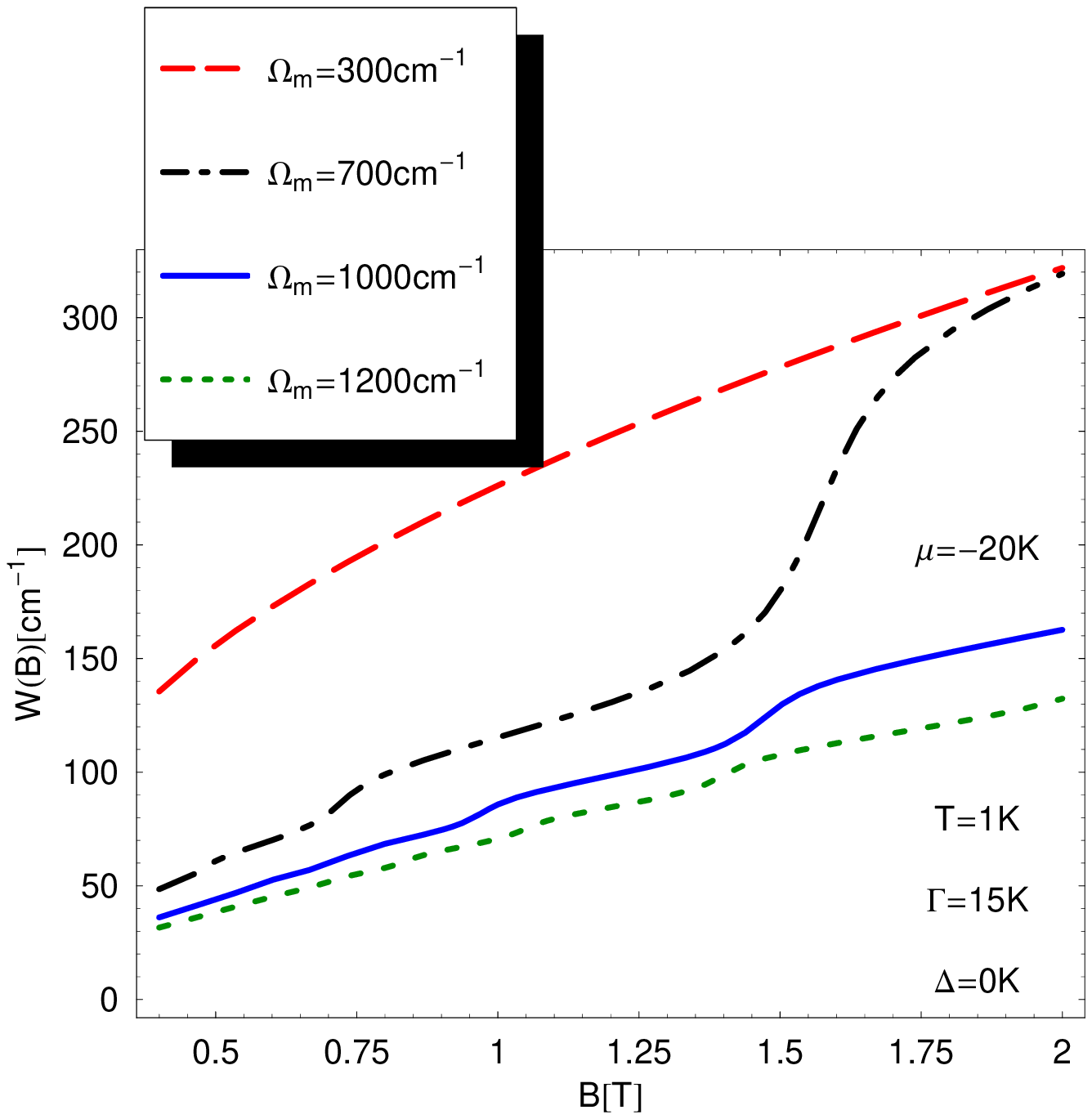}
} \caption{(Color online) Partial Hall spectral weight $W(\Omega_m)$
as a function of field $B$ at $\mu=-20\, \mbox{K}$, temperature $T=
1\, \mbox{K}$, scattering rate $\Gamma =15 \mbox{K}$ for four values
of $\Omega_m$. Long dashed (red) $\Omega_m = 300 \, \mbox{cm}^{-1}$,
dash-dotted (black) $\Omega_m = 700 \, \mbox{cm}^{-1}$, solid (blue)
$\Omega_m = 1000 \, \mbox{cm}^{-1}$, short dashed (green) $\Omega_m
= 1200 \, \mbox{cm}^{-1}$.
 } \label{fig:2}
\end{figure}

Our theory predicts not only the position in energy of the various
lines as well as the shape of the $\mbox{Re}\, t_H(\Omega)$ at small
photon energy, but also provides values for the optical spectral
weight under the various features seen in Fig.~\ref{fig:1}.
Information on this spectral weight is conveniently presented in
terms of the partial sum rule of Eq.~(\ref{W_xy}) for typical values
of the cutoff $\Omega_m$ as a function of the magnetic field. In
Fig.~\ref{fig:2} we show numerical results for the Hall angle
spectral weight $W(\Omega_m)$ in $\mbox{cm}^{-1}$ as a function of
the value of the external magnetic field $B$ for four values of the
cutoff $\Omega_m$ at fixed value of $\mu = -20 \, \mbox{K}$, $T= 1\,
\mbox{K}$ and $\Gamma = 15\, \mbox{K}$.  For the long dashed (red)
curve with the cutoff $\Omega_m = 300 \, \mbox{cm}^{-1}$ only the
frequency region in the long dashed (red) curve of Fig.~\ref{fig:1}
which falls below the first peak (negative in this case) in
$\mbox{Re}\, t_H(\Omega)$ is integrated in the weight. While in
Fig.~\ref{fig:1} the field $B = 0.4\, \mbox{T}$, the peaks in
$\mbox{Re}\, t_H(\Omega)$ move to higher energies as $B$ increases
and so for all values of $B$ used in the figure only the Drude
region around small $\Omega$ of the curve is integrated over. Using
the formula (\ref{t_H-Drude}) as a rough approximation we expect in
this case $W(\Omega_m)$ to scale as the square root of $B$ coming
from the $M_1$ factor. This is verified to good accuracy in the
numerical work. When $\Omega_m$ increases, as is the case for the
other three curves of Fig.~\ref{fig:2}, the peaks in $\mbox{Re}\,
t_H(\Omega)$, which are negative for $\mu = - 20 \, \mbox{K}$, start
entering the integral and this reduces the value of the optical sum.
However, we have no simple approximate analytic formula which might
capture the essence of the situation in this case, so we must rely
on the numerical work. It is also clear that for a given value of
$B$, the reduction in the optical sum caused by the negative peaks
in $\mbox{Re}\, t_H(\Omega)$ will be less as $B$ increases, because
the peaks move to higher energies. In fact the upward steps, seen
most clearly in the dash-dotted (black) curve of Fig.~\ref{fig:2}
with the cutoff $\Omega_m = 700 \, \mbox{cm}^{-1}$, correspond to
values of $B$ for which a peak in $\mbox{Re}\, t_H(\Omega)$  is
starting to fall outside the integration range. The first step
occurs around $B \sim 0.75 \,\mbox{T}$ when the second negative peak
in the long dashed curve of Fig.~\ref{fig:1} is moving through
$700\, \mbox{cm}^{-1}$, while for  $B \sim 1.6 \,\mbox{T}$ it is the
first peak in $\mbox{Re}\, t_H(\Omega)$ which is involved. This peak
is larger in absolute value and so this second step is larger. Note
also that in this case the dash-dotted (black) curve merges with the
long dashed (red) curve for the smaller cutoff as it must, since in
both instances only the Drude-like contribution at small $\Omega$ is
relevant to the integral.

Another feature of the curves in Fig.~\ref{fig:2} is worth
commenting on. We note that the short dashed (green) curve for
$\Omega_m = 1200 \, \mbox{cm}^{-1}$ becomes linear in $B$ at small
$B$ in contrast to the long dashed (red) curve which, as we saw,
went like $\sqrt{B}$. We can get some understanding of this
crossover which corresponds to the case when many peaks are involved
in the integration, by considering the $B \to 0$ limit. In
Ref.~\onlinecite{Gusynin2007JPCM} we also obtained the expressions
(13) for $\sigma_{xx}$ and (14) for $\sigma_{xy}$ which for
$\Delta=0$ read
\begin{equation}
\label{sigma_xx-lowB}
\begin{split}
\sigma_{xx}(\Omega)= - &\frac{2ie^2(\Omega+2i\Gamma)}{h}
\left[\frac{1}{(\Omega+2i\Gamma)^2} \int_{0}^{\infty} d \omega
|\omega|\left(\frac{\partial n_F(\omega)}{\partial
\omega}-\frac{\partial n_F(-\omega)}{\partial
\omega} \right) \right.\\
& -\left.\int_{0}^{\infty} d\omega
\frac{n_F(-\omega)-n_F(\omega)}{(\Omega+2i\Gamma)^2-4\omega^2}\right]
\end{split}
\end{equation}
and
\begin{equation} \label{sigma_xy-lowB}
\begin{split}
\sigma_{xy}(\Omega)=& \frac{e^2v_F^2eB}{\pi c}\int_{0}^{\infty}
d\omega \left(\frac{\partial n_F(\omega)}{\partial \omega}
+\frac{\partial
n_F(-\omega)}{\partial \omega}\right) \\
& \times \left[-\frac{1}{(\Omega+2i\Gamma)^2}+
\frac{1}{4\omega^2-(\Omega+2i\Gamma)^2}\right].
\end{split}
\end{equation}
For   $T=0$ we get
\begin{eqnarray}
\sigma_{xx}(\Omega)&=&\frac{2ie^{2}}{h}\left[\frac{|\mu|}{\Omega+2i\Gamma}+\frac{1}{4}
\ln\frac{2|\mu|-(\Omega+2i\Gamma)}{2|\mu|+(\Omega+2i\Gamma)}\right],\\
\sigma_{xy}(\Omega)&=&\frac{e^{2}v_{F}^{2}eB}{\pi c}\mbox{sgn}\mu\left[\frac{1}{(\Omega+2i\Gamma)^{2}}+
\frac{1}{(\Omega+2i\Gamma)^{2}-4\mu^{2}}\right].
\end{eqnarray}
Hence for $|\mu| \gg \Omega,\Gamma$ we obtain
\begin{equation}
\label{t_H-Drude-B=0} t_H(\Omega) = -\frac{\hbar v_F^2 eB }{c \mu}
\frac{1}{2\Gamma - i\Omega} = - \frac{\hbar eB \, \mbox{sgn}\,
\mu}{c m_c}\frac{1}{2\Gamma - i\Omega},
\end{equation}
where in the second equality we introduced the cyclotron mass $m_c$.
This form is to be contrasted to that obtained in
Eq.~(\ref{t_H-Drude}) for large magnetic field $B$, where
$t_H(\Omega)$ is proportional to $\sqrt{B}$ rather than to $B$ as in
Eq.~(\ref{t_H-Drude-B=0}). Except for a sign change this result is
of the same form as in Eq.~(22) of Ref.~\onlinecite{Drew1997PRL}
with $m_c$ playing the role of mass which in graphene varies as the
square root of the carrier imbalance
$|\rho|$\cite{Geim2005Nature,Kim2005Nature}. Here we emphasize the
linear dependence on magnetic field $B$ in Eq.~(\ref{t_H-Drude-B=0})
as well as in the full sum rule (\ref{sum-rule-Hall}) with
$\omega_H$ given by Eq.~(\ref{omega_H-final-K}).

\begin{figure}[h]
\centering{
\includegraphics[width=8cm]{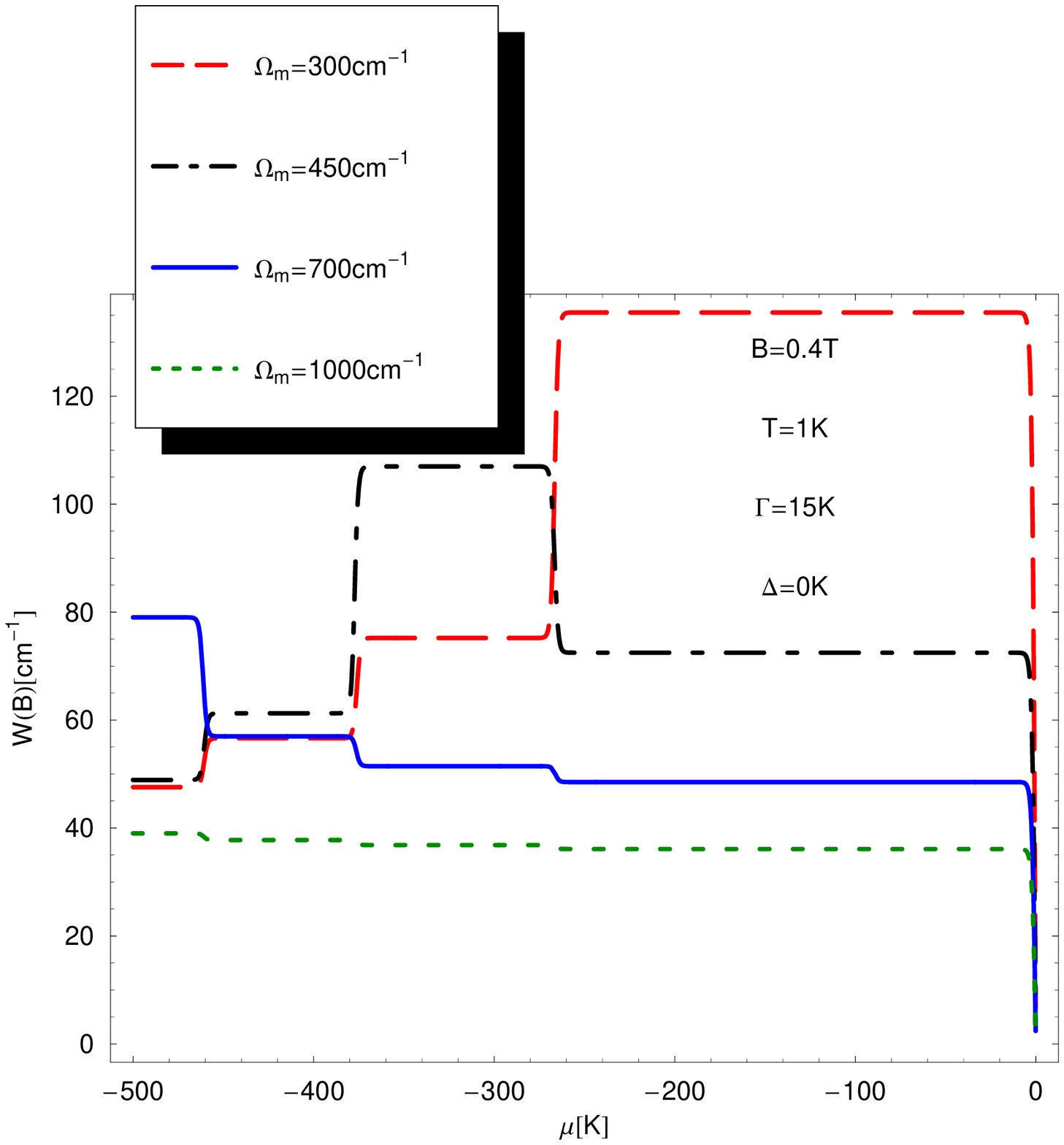}
} \caption{(Color online) Partial Hall spectral weight $W(\Omega_m)$
as a function of $\mu$ at  $B=0.4\, \mbox{K}$, temperature $T= 1\,
\mbox{K}$, scattering rate $\Gamma =15 \mbox{K}$ for four values of
$\Omega_m$.  Long dashed (red) $\Omega_m = 300 \, \mbox{cm}^{-1}$,
dash-dotted (black) $\Omega_m = 450 \, \mbox{cm}^{-1}$, solid (blue)
$\Omega_m = 700 \, \mbox{cm}^{-1}$, short dashed (green) $\Omega_m =
1000 \, \mbox{cm}^{-1}$.} \label{fig:3}
\end{figure}

An experimental configuration which has already been used by Li {\em
et al.} in Ref.~\onlinecite{Basov-organic} is incorporating the
specimen into a field-effect
device\cite{Novoselov2004Science,Geim2005Nature,Kim2005Nature}. In
this case the chemical potential $\mu$ is easily changed by varying
the gate voltage. In Fig.~\ref{fig:3} we show numerical results for
$W(\Omega_m)$ in $\mbox{cm}^{-1}$ as a function of chemical
potential $\mu$ in $\mbox{K}$ for four values of the cutoff
$\Omega_m$. The magnetic field $B$ has been set to $0.4 \,
\mbox{T}$, $T= 1\, \mbox{K}$, and $\Gamma = 15\, \mbox{K}$.  We see
steps occurring in these curves as the chemical potential $\mu$
crosses the Landau level energies -- viz., $M_1 \simeq 265\,
\mbox{K}$, $M_2 \simeq 375\, \mbox{K}$, $M_3 \simeq 460\, \mbox{K}$
and $M_4 \simeq 531\, \mbox{K}$, respectively. Between successive
sharp rises, $W(\Omega_m)$ stays nearly constant quite independent
of $\mu$. The long dashed (red) curve has the smallest cutoff equal
to $300 \, \mbox{cm}^{-1}$ which falls below the first peak in the
long dashed (red) curve of Fig.~\ref{fig:1} for $\mu = - 20 \,
\mbox{K}$ and its value $\sim 135 \, \mbox{cm}^{-1}$ corresponds to
the area under the Drude part of the curve (in Fig.~\ref{fig:1}).
Remaining with $\Omega_{m} = 300 \, \mbox{cm}^{-1}$ as $|\mu|$ is
increased beyond $M_1$ but is less than $M_2$, the value of
$W(\Omega_m)$ decreases. This now corresponds to the dash-dotted
(black) curve of Fig.~\ref{fig:1} which has a smaller Drude-like
piece than does the long dashed (red) curve and the drop is close to
a factor of $0.56$ while our simplified but analytic formula
(\ref{t_H-Drude-N}) predicts $0.71$ (which is better than can be
expected as it is valid only for $N\gg 1$ and we are using it
outside the range of validity). The dash-dotted (black) curve has
$\Omega_m = 450 \, \mbox{cm}^{-1}$ which is chosen to fall above the
first (negative) peak in the long dashed (red) curve of
Fig.~\ref{fig:1}. For $|\mu| < M_1$ the dash-dotted curve of
Fig.~\ref{fig:1} applies, and to get $W(\Omega_m)$, we are
integrating through the first negative peak in addition to the
Drude-like peak centered at $\Omega=0$. This reduces its value to
about half the value it had for the lower cutoff $\Omega_{m} = 300
\, \mbox{cm}^{-1}$. In the region $\mu \in ]M_1,M_2[$, however, the
dash-dotted curve of Fig.~\ref{fig:1} applies and we are integrating
over the first positive peak in $\mbox{Re}\, t_H(\Omega)$ as well as
through the Drude-like region and so the value of the weight
$W(\Omega_m)$ is now increased. After this it decreases again as
more negative peaks are integrated over. The other two curves can be
understood as well with arguments similar to those used for the
first two cutoffs.

\begin{figure}[h]
\centering{
\includegraphics[width=8cm]{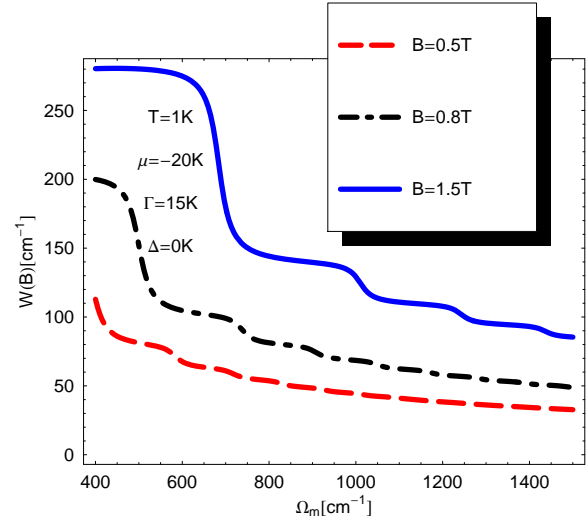}
} \caption{(Color online) Partial Hall spectral weight $W(\Omega_m)$
as a function of the cutoff $\Omega_m$  at  $\mu = -20\,\mbox{K}$,
$T= 1\, \mbox{K}$, scattering rate $\Gamma =15 \mbox{K}$ for three
values of $B$.  Long dashed (red) $B = 0.5 \, \mbox{T}$, dash-dotted
(black) $B = 0.8 \, \mbox{T}$, solid (blue) $B = 1.5 \, \mbox{T}$.}
\label{fig:4}
\end{figure}

Finally in Fig.~\ref{fig:4} we present results for $W(\Omega_m)$ as
a function of $\Omega_m$ for $T= 1\, \mbox{K}$, $\Gamma = 15\,
\mbox{K}$, and $\mu = -20\, \mbox{K}$ for three values of magnetic
field. Each curve shows steps as the various peaks beyond the
Drude-like structure in $\mbox{Re}\, t_H(\Omega)$ vs $\Omega$ are
included in the integral defining $W(\Omega_m)$ as $\Omega_m$ is
increased. These steps are most pronounced for the solid (blue)
curve with $B=1.5\, \mbox{T}$. For this field value the peaks in the
long dashed curve of Fig.~\ref{fig:1} will be shifted upwards by a
factor of $\sim 1.94$, so that the first drop corresponds to
integrating over the first negative peak in the equivalent of the
the long dashed (red) curve of Fig.~\ref{fig:1}, while the second
drop corresponds to integrating over the second negative peak etc.
The drops in the other curves can be similarly traced and fall at
smaller values of $\Omega_m$ because of the $\sqrt{B}$ scaling of
the peak positions in $\mbox{Re}\, t_H(\Omega)$ vs $\Omega$. To end
we note that $W(\Omega_m)$ could also show upward rather than
downward steps if we had chosen a larger value of $|\mu|$ between
$M_1$ and $M_2$. In this case as an example, the first peak in
$\mbox{Re}\, t_H(\Omega)$ is positive, so it adds to the low
$\Omega$ Drude-like contribution.

\section{Conclusion}
\label{sec:concl}

We have studied the modification of the usual optical sum rules
brought about by the particular band structure of graphene --
namely, its compensated semimetal aspects and the Dirac nature of
its quasiparticles. The existence of two bands within the same
Brillouin zone leads to interband as well as intraband optical
transitions both of which enter the optical sums.

We considered both the usual $f$-sum rule on the real part of the
longitudinal conductivity $\sigma_{xx}(\Omega)$ and an equivalent
sum rule recently introduced by Drew and Coleman \cite{Drew1997PRL}
involving the Hall conductivity. From a formal point of view we find
that care must be used in introducing the electromagnetic vector
potential. It should be introduced into the initial Hamiltonian
through a Peierls phase factor rather than after the
diagonalization. The $f$-sum rule is shown to be proportional to
$\sim -0.39$ of the kinetic energy per atom rather than the usual
minus one-half value, familiar for a square lattice with
nearest-neighbor hopping only. This difference reflects the
particularity of the honeycomb lattice with nearest-neighbor hopping
between the two distinct $A$ and $B$ sublattices.

For small chemical potential $|\mu|<< T$, the longitudinal sum rule
displays a $T^3$ temperature law rather than conventional $T^2$ law
of free electron theory. Also for $|\mu| \gg T$ we predict a
$|\mu|^3$ dependence on chemical potential  for the deviation from
the $T=\mu=0$ reference case. The first temperature correction in
this case goes as $|\mu|T^2$, recovering the $T^2$ law of free
electron theory. The $|\mu|^3 \varpropto |V_g|^{3/2}$ dependence
should be observable in field-effect devices.

The sum rule for the Hall conductivity involves the real part
$\mbox{Re}\, t_H(\Omega)$ of the Hall angle (\ref{t_H}). Its
integral (\ref{sum-rule-Hall}) is equal to the Hall frequency
$\omega_H$, and in the free electron metals $\omega_H$ coincides
with the cyclotron frequency. For graphene we find instead that
$\omega_H$ is given by Eq.~(\ref{omega_H-final-K}). Although this
equation can be formally written in the form (\ref{omega_H-filled})
which involves the ``relativistic'' cyclotron mass $m_c =
|\mu|/v_F^2$, expression (\ref{omega_H-filled}) is different from
the conventional one. There are two modifications. The cyclotron
mass $m_c$ varies as the square root of the carrier density, and
$\omega_H$ in Eq.~(\ref{omega_H-filled}) is also proportional to the
ratio $(\mu/t)^3$.

In our previous paper \cite{Gusynin2007JPCM} we considered partial
spectral weight $W_{xx}(\Omega_m)$ given by Eq.~(\ref{W_xx}) for
$\mbox{Re}\, \sigma_{xx}(\Omega)$ and found steplike structures
corresponding to peaks in $\mbox{Re}\, \sigma_{xx}(\Omega)$. Here we
considered the partial spectral weight $W(\Omega_m;B,\mu)$ given by
Eq.~(\ref{W_xy}) for $\mbox{Re}\, t_H(\Omega)$. Interesting steplike
structures are also found in which the steps can go up or down
depending on the variable used to display the weight
$W(\Omega_m;B,\mu)$. Specifically we analyzed its dependence on the
value of the external magnetic field $B$ and chemical potential for
various fixed values of $\Omega_m$. We also considered the
dependence of $W(\Omega_m;B,\mu)$ on $\Omega_m$ for fixed $\mu$ and
$B$ values. The rich pattern of behavior can  be traced back to the
frequency dependence of $\mbox{Re}\, t_H(\Omega)$ which shows a
Drude-like peak at small $\Omega$ followed by a series of peaks
which fall at some energy between the interband peaks of
$\sigma_{xx}(\Omega)$, where both its real and imaginary parts are
small. Also the peaks in $\mbox{Re}\, t_H(\Omega)$ depend on the
sign and value of chemical potential.

We hope that these specific predictions for the behavior of
$\mbox{Re}\, t_H(\Omega)$, the corresponding partial spectral weight
$W(\Omega_m;B,\mu)$, and the sum rules can be verified
experimentally.

\section*{Acknowledgments}

The work of V.P.G. was supported by the SCOPES-project IB7320-110848
of the Swiss NSF and by Ukrainian State Foundation for Fundamental
Research. J.P.C. and S.G.Sh. were supported by the Natural Science
and Engineering Research Council of Canada (NSERC) and by the
Canadian Institute for Advanced Research (CIAR).

\appendix
\section{Diamagnetic term}
\label{sec:A}

Substituting the Fourier transform (\ref{Fourier}) into
Eq.~(\ref{diamagnetic-term}) and summing over the lattice sites, we
obtain
\begin{equation} \label{tau-Fourier}
\begin{split}
\sum_{\mathbf{n}}\tau_{\alpha \beta}(\mathbf{n})& =
-\frac{te^2}{\hbar^2}\sum_{\pmb{\delta},\sigma}\delta_\alpha
\delta_\beta\int_{BZ}\frac{d^2{\mathbf{k}}}{(2\pi)^2}
\left[a_{\sigma}({\mathbf{k}})^\dagger e^{i \mathbf{k}\pmb{\delta}
}b_{\sigma}({\mathbf{k}})+\mbox{h.c.}\right]\\
&=
\frac{e^2}{\hbar^2}\sum_{\sigma}\int_{BZ}\frac{d^2{\mathbf{k}}}{(2\pi)^2}\left[
\frac{\partial^2 \phi(\mathbf{k})}{\partial k_\alpha\partial
k_\beta} \Upsilon_{\sigma}^\dagger (\mathbf{k}) \sigma_+
\Upsilon_{\sigma} (\mathbf{k}) + \frac{\partial^2
\phi^\ast(\mathbf{k}) }{\partial k_\alpha\partial k_\beta}
\Upsilon_{\sigma}^\dagger (\mathbf{k}) \sigma_- \Upsilon_{\sigma}
(\mathbf{k})\right].
\end{split}
\end{equation}
Accordingly the thermal average $\langle \tau_{\alpha \beta}\rangle$
is easily expressed in terms of the GF (\ref{GF-basic}) as
\begin{equation}
\begin{split}
\label{tau-off1} \langle \tau_{\alpha \beta}\rangle = & \frac{2
Ve^2}{\hbar^2} \int_{BZ} \frac{d^2{\mathbf{k}}}{(2\pi)^2}T
\sum_{n=-\infty}^\infty
e^{-i \omega_n \tau} \\
& \times \left[ \mbox{tr}[\sigma_+ G (i \omega_n,\mathbf{k})]
\frac{\partial^2 \phi(\mathbf{k})}{\partial k_\alpha\partial
k_\beta}+ \mbox{tr}[\sigma_- G (i \omega_n,\mathbf{k})]
\frac{\partial^2 \phi^\ast(\mathbf{k})}{\partial k_\alpha\partial
k_\beta}  \right], \qquad \tau \to 0.
\end{split}
\end{equation}
Calculating the trace we obtain for the diagonal component of the
stress tensor $\langle \tau_{\alpha \beta}\rangle = \langle
\tau_{\alpha \alpha}\rangle \delta_{\alpha \beta}$ that
\begin{equation}
\label{tau-off2} \frac{\langle\tau_{\alpha
\alpha}\rangle}{V}=-\frac{2e^2T}{\hbar^2}\sum\limits_{n=-\infty}^\infty\int_{BZ}\frac{d^2{\bf
k}}{(2\pi)^2}
\frac{1}{(\omega_n-i\mu)^2+\epsilon^2(\mathbf{k})}\left(\phi(\mathbf{k})\frac{\partial^2}
{\partial k_\alpha^2}\phi^\ast(\mathbf{k})+\mbox{c.c.}\right).
\end{equation}
The sum over the Matsubara frequencies converges irrespectively of
infinitesimally small $\tau$, and we find
\begin{equation}
 \label{tau-average}
\frac{\langle\tau_{\alpha \alpha}\rangle}{V}=
\frac{e^2}{\hbar^2}\int_{BZ}\frac{d^2{\bf k}}{(2\pi)^2}
\frac{1}{2\epsilon(\mathbf{k})}\left[n_F\left(\epsilon(\mathbf{k})\right)-n_F\left(-\epsilon(\mathbf{k})\right)\right]
\left(\phi(\mathbf{k})\frac{\partial^2} {\partial
k_\alpha^2}\phi^\ast(\mathbf{k})+\mbox{c.c.}\right).
\end{equation}
Writing $\phi(\mathbf{k})=
\epsilon(\mathbf{k})e^{i\varphi(\mathbf{k})}$ we can recast the last
equation in the form (\ref{sum-rule-phase}). The derivatives in the
brackets in Eq.~(\ref{tau-average}) are easily calculated using the
explicit expression for $\phi(\mathbf{k})$:
\begin{equation} \label{derivative} \phi(\mathbf{k})\frac{\partial^2} {\partial
k_\alpha^2} \phi^\ast(\mathbf{k})+\mbox{c.c.}=
\frac{1}{2}\sum_{\alpha=1,2}\left(\phi(\mathbf{k})\frac{\partial^2}
{\partial k_\alpha^2} \phi^\ast(\mathbf{k})+\mbox{c.c.}\right)
=-\frac{a^2}{3} \epsilon^2(\mathbf{k}).
\end{equation}
Substituting Eq.~(\ref{derivative}) into Eq.~(\ref{tau-average}) we
arrive at the final expression (\ref{sum-rule-main}).

\section{Hall term}
\label{sec:B}

Here we calculate the commutator $I_{x,y} = \langle [J_x(t=0),
J_y(t=0)]\rangle$ of the paramagnetic currents $J_{x}(t=0)$ and
$J_{y}(t=0)$ that are defined in Eq.~(\ref{current_t=0}). Since this
commutator is nonzero only in the presence of a magnetic field, the
corresponding current density operator $j_a(\mathbf{n})$ has to be
taken for a finite vector potential $\mathbf{A}$:
\begin{equation}
\label{current-param-A} j_a(\mathbf{n})=  -\frac{ite}{\hbar}
\sum_{\pmb{\delta},\sigma}\delta_a\left[a_{\mathbf{n},\sigma}^\dagger
T_{\pmb{\delta}} b_{\mathbf{n},\sigma} -
b_{\mathbf{n},\sigma}^\dagger T_{-\pmb{\delta}}
a_{\mathbf{n},\sigma}\right],
\end{equation}
where the operator
\begin{equation}
\label{T} T_{\pmb{\delta}}= e^{i \pmb{\delta}(\mathbf{p} + e/c
\mathbf{A})/\hbar} .
\end{equation}
Using the anticommutation relations for $a_{\mathbf{n},\sigma}$ and
$b_{\mathbf{n},\sigma}$ operators we calculate the commutator
\begin{equation}
\label{commutator-nm} [j_x (\mathbf{n}), j_y (\mathbf{m})] =
\frac{e^2 t^2\delta_{nm}}{\hbar^2} \sum_{\pmb{\delta},
\pmb{\delta}^\prime, \sigma} \delta_x \delta_y^\prime
\left\{a_{\mathbf{n},\sigma}^\dagger\left(T_{\pmb{\delta}}T_{-\pmb{\delta}^\prime}-
T_{\pmb{\delta}^\prime}T_{-\pmb{\delta}}\right)a_{\mathbf{n},\sigma}
+b_{\mathbf{n},\sigma}^\dagger\left(T_{-\pmb{\delta}}T_{\pmb{\delta}^\prime}-
T_{-\pmb{\delta}^\prime}T_{\pmb{\delta}}\right)b_{\mathbf{n},\sigma}\right\}.
\end{equation}
Provided that the magnetic flux per unit cell is far less than a
flux quantum $h/e$, we may expand the operator  $T_{\pmb{\delta}}$
\begin{equation}
\label{T-expand} T_{\pmb{\delta}} \approx 1 + i
\frac{\delta_a}{\hbar} \left(p_a + \frac{e}{c}A_a\right).
\end{equation}
Then using the commutator
\begin{equation}
[D_a,D_b] \equiv \left[p_a + \frac{e}{c}A_a, p_b +
\frac{e}{c}A_b\right] = - i \frac{e\hbar}{c} B \epsilon_{ab}, \qquad
a,b=1,2
\end{equation}
we obtain
\begin{equation}
\label{T-identity} T_{\pmb{\delta}}T_{-\pmb{\delta}^\prime}-
T_{\pmb{\delta}^\prime}T_{-\pmb{\delta}} =
\frac{1}{\hbar^2}[\delta_a \delta_b^\prime D_a D_b - \delta_a^\prime
\delta_b D_a D_b] = \frac{\delta_a \delta_b^\prime}{\hbar^2} (-i)
\frac{e\hbar}{c} B \epsilon_{ab},
\end{equation}
where the sum over the dummies $a,b$ is implied. Now the sum over
$\pmb{\delta}, \pmb{\delta}^\prime$ in Eq.~(\ref{commutator-nm}) can
be evaluated as follows
\begin{equation}
\label{sum-delta} \sum_{\pmb{\delta}, \pmb{\delta}^\prime} \delta_x
\delta_y^\prime \delta_\gamma \delta_\kappa^\prime \epsilon_{\gamma
\kappa} =
 \epsilon_{\gamma \kappa} \sum_{\pmb{\delta}} \delta_x
 \delta_\gamma \sum_{\pmb{\delta}^\prime}
\delta_y^\prime  \delta_\kappa^\prime = \epsilon_{xy} \frac{a^4}{4},
\end{equation}
where in the second equality we used the relation
\begin{equation}
\label{completness} \sum_{\lambda=1,2,3}(\delta_\lambda)_\alpha
(\delta_\lambda)_\beta=\frac{a^2}{2}\delta_{\alpha \beta}.
\end{equation}
Substituting the commutator (\ref{commutator-nm}) into the
expression $I_{x,y} = \sum_{\mathbf{n},\mathbf{m}} \langle
[j_x(t=0,\mathbf{n}), j_y(t=0,\mathbf{m})]\rangle$ and utilizing
Eqs.~(\ref{T-identity}) and (\ref{sum-delta}) we arrive at the
representation
\begin{equation}
\label{current-commutator}
I_{x,y}  = - i \frac{e^2 a^4 t^2 \hbar e
B}{4 \hbar^4 c} \epsilon_{xy} \sum_{\mathbf{n},\sigma} \langle
(a_{\mathbf{n},\sigma}^\dagger a_{\mathbf{n},\sigma}
+b_{\mathbf{n},\sigma}^\dagger b_{\mathbf{n},\sigma})\rangle .
\end{equation}
Thus to complete the calculation of $I_{x,y}$ we have to find the
thermal average
\begin{equation}
\tilde{I} \equiv \sum_{\mathbf{n},\sigma} \langle(
a_{\mathbf{n},\sigma}^\dagger a_{\mathbf{n},\sigma} +
b_{\mathbf{n},\sigma}^\dagger b_{\mathbf{n},\sigma}) \rangle.
\end{equation}
As in Appendix~\ref{sec:A}, $\tilde{I}$  is expressed in terms of
the GF (\ref{GF-basic}) as
\begin{equation}
\label{I-tilde} \tilde{I}  = \sum_{\mathbf{n}} \sum_\sigma
\mbox{tr}[\hat{I}G_\sigma(\tau,\mathbf{0})] = 2 V T \sum_n \int_{BZ}
\frac{d^2{\mathbf{k}}}{(2\pi)^2} [G_{11}(i
\omega_n,\mathbf{k})+G_{22}(i \omega_n,\mathbf{k})] e^{-i \omega_n
\tau},
\end{equation}
where
\begin{equation}
\label{G-expanded} G_{11}(i \omega_n,\mathbf{k}) = G_{22}(i
\omega_n,\mathbf{k}) = \frac{1}{2}\left[\frac{1}{i\omega_n + \mu -
\epsilon(\mathbf{k})} + \frac{1}{i\omega_n + \mu +
\epsilon(\mathbf{k})} \right]
\end{equation}
are the diagonal in the sublattice space components of the GF
(\ref{GF-basic}). In contrast to the convergent Matsubara sums
(\ref{tau-off1}) and (\ref{tau-off2}) in the off-diagonal components
of the GF (\ref{GF-basic}), regularization of the Matsubara sum
(\ref{I-tilde}) by the factor $e^{-i \omega_n \tau}$ is crucial,
because it would diverge otherwise. The factor $e^{-i \omega_n
\tau}$ regularizes the sum over Matsubara frequencies differently
depending on the sign of $\tau$:
\begin{equation}
\label{prescription} \lim_{\tau \to \pm 0} T \sum\frac{e^{-i \tau
\omega_n}}{i\omega_n + \mu - \omega} = - \frac{1}{2}[1-2
n_F(\omega)] - \frac{\mbox{sgn} \tau}{2} = \left\{
                      \begin{array}{lr}
                      n_{F}(\omega),  & \tau \to -0, \\
                         n_{F}(\omega)-1, &  \tau \to +0. \\
                      \end{array}
                    \right.
\end{equation}
Normally to count the number of particles \cite{Abrikosov.book} one
takes the limit $\tau \to -0$. One can check, however, that such a
prescription leads to a contradiction when calculating the integral
(\ref{I-tilde}), because it results in $\tilde{I}$ which is not odd
in $\mu$, while the LHS of Eq.~(\ref{sum-rule-Hall}) is odd in
$\mu$. This contradiction is resolved if one takes into account that
the first term of the GF (\ref{G-expanded}) describes electrons, so
that the prescription $\tau \to -0$ has to be used, while the second
term of the GF (\ref{G-expanded}) describes holes and the
prescription $\tau \to +0$ is necessary. Then we obtain
\begin{equation}
\label{tilde-I-final}
 \tilde{I}  = 2 V  \int_{BZ} \frac{d^2{\mathbf{k}}}{(2\pi)^2}
[n_F(\epsilon(\mathbf{k})) + n_F(-\epsilon(\mathbf{k})) -1],
\end{equation}
which is odd in $\mu$. The quantity (\ref{tilde-I-final}) describes
the carrier imbalance $\rho = n_e - n_h$, where $n_e$ and $n_h$ are
the densities of electrons and holes, respectively. It was
considered in the Dirac approximation in Appendix~C of
Ref.~\onlinecite{Gusynin2006PRB}, and  here we use two expressions
(\ref{rho}) and (\ref{rho-gap}) for $\rho$.

\newpage

\end{document}